# A Control-Theoretic Approach to Dynamic Payment Routing for Success Rate Optimization

Aniket Agrawal
JUSPAY
Bangalore,India
aniket.agrawal.c2023@iitbombay.org

Harsharanga Patil
JUSPAY
Bangalore,India
harsharanga@juspay.in

## ABSTRACT

This paper introduces a **control-theoretic framework** for dynamic payment routing, implemented within JUSPAY's Payment Orchestrator to maximize transaction success rate. The routing system is modeled as a **closed-loop feedback controller** continuously sensing gateway [3] performance, computing corrective actions, and dynamically routes transactions across gateway to ensure operational resilience.

The system leverages concepts from **control theory**, **reinforcement learning**, and **multi-armed bandit optimization** to achieve both short-term responsiveness and long-term stability. Rather than relying on explicit **PID** regulation, the framework applies **generalized feedback-based adaptation**, ensuring that corrective actions remain proportional to observed performance deviations and the computed gateway score gradually converges toward the success rate [2].

This hybrid approach unifies **control theory** and **adaptive decision systems**, enabling self-regulating transaction routing that dampens instability, and improves reliability. Live production results show an improvement of up to **1.15%** in **success rate** over traditional rule-based routing, demonstrating the effectiveness of feedback-based control in payment systems.

# 1. System Architecture

## 1.1 Overview

The system architecture for dynamic payment routing can be interpreted through the lens of **control theory** as a **closed-loop feedback system**. The **Decision Engine** functions as the controller, continuously regulating gateway selection based on real-time performance signals, while the **Feedback Loop** acts as the sensing mechanism that observes transaction outcomes and updates gateway scores.

Together, they form an adaptive control cycle that filters, orders, initiates, and corrects routing decisions dynamically in response to changing gateway conditions.

## 1.2 System Components

### 1.2.1 Decision Engine (Pre-Transaction)
This is the core component of the application, responsible for filtering, ordering and detecting downtimes to finally select the optimal gateway for each transaction. It operates as follows:

- **Eligibility Check**: Filters the gateways based on merchant-defined eligibility check, such as payment instrument, card bin, enablement and other custom configurations specific to the merchant's requirements
- **Dynamic Gateway Ordering**: Orders the eligible gateways based on their **Success Rate Score (SR Score)**, computed using the success rates of recent transactions.
- **Downtime Detection** [8]: Monitors gateway performance based on merchant-defined metrics and dynamically reorders gateways that demonstrate poor performance metrics in real-time.
- **Cascading Retries** [9]**:** The gateway at the top of the list is selected for the transaction. If the initiation call to this gateway fails, the system retries with subsequent gateways in the list, continuing this process for the number of retries configured by the merchant, and sends the feedback for each failed gateway as FAILURE to update their scores.

### 1.2.2 Configurations & Dimensions
The system maintains multiple SR and Health scores for each gateway, reflecting performance across various dimensions. A dimension is defined by a combination of fields such as **MERCHANT_ID, PLATFORM, PAYMENT_INSTRUMENT, NETWORK,** etc. These can be extended to additional fields based on merchant requirements and internal analysis. The optimal parameters, defined in the algorithm, to compute the scores are auto-configured across all dimensions and remain the same for all gateways, ensuring fairness and eliminating any bias towards specific gateways.

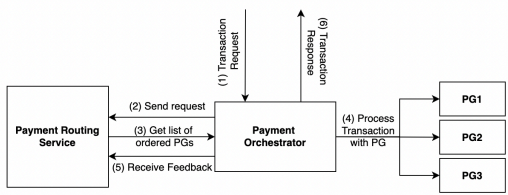

Figure 1: System Diagram for Payment Routing Flow

### 1.2.3 Feedback Loop (Post-Transaction)

This component operates asynchronously to update scores based on **REWARD** and **PENALIZE** feedback generated for each initiated transaction. The following feedback mechanisms are implemented:

- **SR & Health Score REWARD**: Sent when a transaction succeeds within the timeout of TP 99.9% (~3 minutes).
- **SR Score PENALIZE**: Sent when a transaction fails within the timeout of TP 99% (~90 seconds). If no feedback is provided within this timeframe, a default penalize is applied.
- **Health Score PENALIZE**: Sent whenever a transaction is initiated to a gateway.

### 1.2.4 Experimentation Platform

The system integrates an external experimentation platform to evaluate diverse configurations (unique set of parameters). The platform dynamically supplies experimental configurations to the decision engine, ensuring equal transaction distribution for performance comparison. Key features include:

- **Isolated Scores**: Maintains separate scores for each configuration to ensure independent evaluation.
- **A/B Testing** [10]: Splits transactions equally across configurations to compare performance.
- **Dimension Extension**: Supports the addition of new fields in dimensions to assess their impact on routing performance.

## 1.3 Reliability Measures

To ensure system reliability, the following alerts and monitoring mechanisms are implemented:

- **Metric Monitoring**: Alerts for stagnant scores, continuous downtime exceeding 2 hours, and other anomalies.
- **Code Efficiency**: Enables safe code changes without disruptions.
- **Change Detection**: Identifies merchant or gateway-side changes that impact their success rate.

## 1.4 Summary

This architecture provides a scalable, reliable and dynamic routing system. By combining real-time feedback, robust storage solutions and experimentation capabilities, the system ensures optimal payment gateway selection with minimal additional latency (~5ms for Decision Engine execution) to maximize transaction success rates.

# 2. Dynamic Gateway Ordering

## 2.1 Overview

Dynamic Gateway Ordering represents the control action where, after filtering out ineligible gateways, the system ranks the list of eligible gateways in descending priority for transaction routing. The primary objective is to maximize the success rate ($SR$) [2] of the transactions by leveraging a scoring mechanism based on recent gateway performance.

## 2.2 Gateway Scoring & Ordering

The scoring mechanism is based on a sliding window technique, as illustrated in figure 2, to evaluate the success rates of the last $n$ transactions for each gateway at all dimensions. Once we receive the feedback, we maintain the window like this:

1. Each gateway's transaction status is added to its respective sliding window.
2. The earliest transaction in the window is removed to maintain a fixed window size ($n$).
3. Maintain the count of successful transactions in the window.

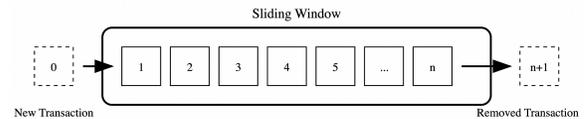

**Figure 2: Illustration of Sliding Window Mechanism**

Based on this count we compute the success rate score in decision engine for eligible gateways:

$$Score = \frac{Successful\ Transactions\ in\ Window}{Window\ Size}$$

The gateways are then ordered in the descending order of their scores. This ensures that, barring any downtimes, the highest-ranked gateway is selected for routing the transaction.

## 2.3 Practical Problems & Modifications

While implementing the dynamic gateway routing system, three practical challenges were identified, and appropriate modifications were introduced to address them

### 2.3.1 Starvation to Single-Gateway
**Problem:** When a gateway becomes the best-performing option, the system tends to route all traffic to it, stopping traffic to other gateways. This creates a starvation problem, as it prevents the system from gathering recent performance data for non-selected gateways.

**Solution:** To address this, the system continuously evaluates all gateways by allocating a small percentage of transactions for each gateway (exploration [14, 16], approx 5-10%). This ensures that the performance of all gateways is monitored and up-to-date.

### 2.3.2 Sudden Changes in Traffic to a Gateway

**Problem:** Gateway SR can fluctuate, leading to scenarios where a gateway's score improves suddenly, causing an abrupt increase in traffic (e.g., from 5% to 85%). This surge inflates the gateway's score, as successful transactions are processed quickly, while failures take longer to register (observed in real data). Meanwhile, the deselected gateway's score deteriorates because failures continue to arrive, but successful transactions stop, compounding the score imbalance.

**Solution:** To prevent bias and ensure fairness across gateways, the system only considers the transactions through exploration for scoring all gateways. This ensures that scores are computed based on same time intervals and transaction volumes across all gateways, avoiding artificial inflation or degradation of scores due to sudden traffic shifts.

### 2.3.3 Maintaining long-term data

**Problem:** Long-term data often becomes less reflective of current conditions and less responsive to recent issues or fluctuations in performance. This is especially problematic in the payments ecosystem where success rates and user intent changes dynamically over time.

**Solution:** Reinforcement Learning (RL) [5, 6] was adopted to address the limitations of AI-based methods by dynamically adapting to real-time performance. Unlike traditional approaches that heavily rely on historical data, RL continuously learns and updates routing decisions to handle rapid changes effectively.

By addressing these challenges, the system ensures robust and reliable gateway selection while maintaining fairness in gateway evaluations.

## 2.4 Mathematical Formulation

Dynamic Gateway Ordering can be formalized as a mathematical framework to optimize gateway selection based on transaction success rates. This section presents the problem modeling and parameter optimization required to achieve the best average success rate ($SR$) across gateways.

### 2.4.1 Problem Modeling

The problem of selecting the best gateway can be mapped to a **Non-stationary Multi-Armed Bandit (MAB)** [12, 13] problem with **Delayed Feedback** [15], where:

- Each gateway is an "arm" with fluctuating success rates and varying latency for success and failure.
- The **explore-exploit** method [14, 16] balances two goals:
    - **Exploration**: Continuous evaluation of all gateways by sending a small percentage of traffic to ensure up-to-date performance data.
    - **Exploitation**: Routing the majority of traffic to the best-performing gateway to maximize the overall success rate.

The system introduces key parameters to control this behavior:

- **Window Size** ($n$): Defines the number of transactions considered for computing a gateway's success rate. This parameter affects the system's responsiveness and stability.
- **Exploration Factor** ($e$): Determines the percentage of traffic allocated to exploration, ensuring fairness and avoiding starvation of lower-ranked gateways.

Success rates are influenced by user intent, which varies over time, making it inaccurate to combine data from different periods. To ensure reliable evaluation, the window should contain only recent data, typically from the last 2 hours.

### 2.4.2 Parameter Optimization

To achieve the best performance, the parameters **n** and **e** are derived using long-term data.

**Example Simulation**
Consider two gateways with fluctuating success rates:

- Gateway 1 ($GW_1$): ~80%
- Gateway 2 ($GW_2$): ~81%

If transactions are distributed randomly between the gateways, the average SR would be **80.5%**. However, since gateway success rates fluctuate over time, the optimized parameters ensure adaptability, allowing the system to outperform rule-based priority selection by achieving a higher average SR of **~80.655%**, even when a previously lower-performing gateway (e.g., $GW_1$) becomes better.

**Modeling Gateway Score**:

The success rate score based on last $n$ transactions for $GW_1$ and $GW_2$, represented as random variables $X_1$ and $X_2$ respectively, follow binomial distributions [4] under the assumption of independent trials. This can be further approximated as normal distributions [4, 7].

$X_1 \sim N(\mu_1 = 0.8, \sigma_1^2 = \frac{0.8 \cdot 0.2}{n})$

$X_2 \sim N(\mu_2 = 0.81, \sigma_2^2 = \frac{0.81 \cdot 0.19}{n})$

The Difference $D = X_2 - X_1$ follows:
$D \sim N(\mu_D = \mu_2 - \mu_1, \sigma_D^2 = \sigma_1^2 + \sigma_2^2)$
$\mu_D = 0.81 - 0.8 = 0.01$
$\sigma_D^2 = \frac{0.8 \cdot 0.2}{n} + \frac{0.81 \cdot 0.19}{n} = \frac{0.3139}{n}$

The probability of $GW_2$ outperforming $GW_1$ is
$P(X_2 > X_1) = P(D > 0) = P(Z > (0 - \mu_D)/\sigma_D)$
where, $Z \sim N(0, 1)$, substituting $\mu_D$ and $\sigma_D$ gives:

$P(X_2 > X_1) = P(Z > -\frac{0.01}{\sqrt{0.3139/n}})$
$= P(Z > -\sqrt{\frac{n}{3139}})$

**Optimal Exploration Factor**:

Now, for Exploration Factor ($e$) (ranging b/w 0 to 1) and transaction rate of 1 transaction per second ($TPS$) for the current dimension, the window size ($n$) is:
$n = e * 2 * 60 * 60 * TPS = 7200 * e$

Thus, the fraction of volume allocated to the best gateway is given by:
$V(e) = e + (1 - 2e) \cdot P(Z > -\sqrt{2.294e})$Í
(exploration) (exploitation)

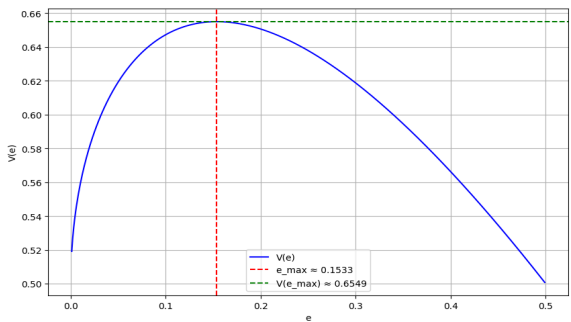

Figure 3: Plot of V(e) vs e

The plot in figure 3 for $V(e)$ vs $e$ gives us the optimal exploration factor & window size.
$e \approx 0.1533, n \approx 1104$
In general, we need to consider m gateways, which follows:

$V(e) = e + (1 - m \cdot e) \cdot \prod_{i=2}^{m} P(Z > -\sqrt{k_i \cdot e})$

where, $k_i$ is dependent on $SR$ of $GW_1$, $SR$ of $GW_i$ & $TPS$

### 2.4.3 Scaling Across Dimensions

This approach is scaled across dimensions for all merchants by configuring optimal parameters based on long-term data of each dimension, including:

- Transaction Per Second ($TPS$)
- Long-Term Gateway Performance
- Number of Available Gateways

By leveraging historical data, the system configures parameters that adapt to varying success rates of gateways, ensuring optimal routing performance and robust gateway selection.

## 3. Downtime Detection

### 3.1 Overview

Downtime detection [8] dynamically identifies and deprioritizes gateways that exhibit sudden performance degradation based on real-time merchant-defined metrics such as latency, failure-rate and error-rate. This approach minimizes business losses by reordering gateway selection to ensure a smooth payment flow, while effectively handling payment system downtimes in both rule-based and dynamic gateway ordering.

### 3.2 Scoring & Downtime Threshold

The downtime detection mechanism assigns a dynamic score to each gateway based on its real-time performance. The score is adjusted using a reward and penalize feedback loop inspired by the **PID** controllers [11]:

- **Penalize**: Reduces the score when a transaction is initiated, ensuring transactions avoid unresponsive gateways.
  $score\_new = score\_old \times (1 - reward\_factor)$
- **Reward**: Increases the score when a transaction is charged, signifying improved performance.
  $score\_new = score\_old + reward\_factor$

If the score drops below the configured threshold, the gateway becomes down. The reward factor ($a$) and threshold ($t$) are computed using long-term data specific to each dimension for a merchant, including metrics such as average success rate ($SR$), average transactions per second ($TPS$), and average latency.

## 3.3 Mathematical Formulation

The score update logic per transaction can be modelled as a function of the average success rate ($SR$):

$$score\_new = (1 - a) \cdot score\_old + a \cdot \frac{SR}{100}$$
$$\quad\quad\quad (Penalize) \quad\quad (Reward * P(Reward))$$

The score's mean and standard deviation can be expressed as:

$$score\_mean = \frac{SR}{100}$$

$$score\_std = \sqrt{\frac{a}{2-a} \cdot \frac{SR}{100} \cdot (1 - \frac{SR}{100})}$$

$$score\_mean - n \cdot score\_std = score\_threshold$$

Here, n is the sigma factor [4] representing the allowable false downtime detection rate, derived using TPS for the dimension.

When the success rate drops from $SR_1$ (average) to $SR_2$ (downtime SR threshold), the score follows this exponential decay function over transaction count ($T$) based on above modelled function:

$$score(T) = (\frac{SR_1 - SR_2}{100}) \cdot e^{-a.T} + \frac{SR_2}{100}$$

To determine the transaction count ($t_c$) needed for the score to fall below the threshold, we derive:

$$\frac{SR_1}{100} - n \cdot score\_std = (\frac{SR_1 - SR_2}{100}) \cdot e^{-a.t_c} + \frac{SR_2}{100}$$

$$t_c = -\frac{1}{a} \ln(1 - \frac{100 \cdot n \cdot score\_std}{SR_1 - SR_2})$$

Substituting $score\_std$ into the equation, we will get:

$$t_c = -\frac{1}{a} \ln(1 - k \cdot \sqrt{\frac{a}{2-a}})$$

where, $k = \frac{n \cdot \sqrt{SR_1(100 - SR_1)}}{(SR_1 - SR_2)}$

To minimize $t_c$, we compute its derivative w.r.t to $a$ & equate it to 0.

$$=> \ln(1 - k \cdot \sqrt{\frac{a}{2-a}}) + \frac{ak}{(2-a) \cdot (\sqrt{a \cdot (2-a)} - ak)} = 0$$

$$=> \ln(1 - k \cdot \sqrt{\frac{a}{2-a}}) + \frac{1}{(2-a) \cdot (\frac{\sqrt{(2-a)/a}}{k} - 1)} = 0$$

$$=> \ln(1 - x) + \frac{1}{(2-a)(\frac{1}{x} - 1)} = 0, \text{ where } x = k \cdot \sqrt{\frac{a}{2-a}}$$

Approximating $2 - a \approx 2$, as $reward\_factor$ for a single transaction is small, we will get:

$$\ln(1 - x) \cdot \frac{(1-x)}{x} + \frac{1}{2} = 0$$

Solving this equation, gives:

$$x \approx 0.715331863 \approx \sqrt{\frac{1}{2}}$$

$$=> k \cdot \sqrt{\frac{a}{2-a}} \approx \sqrt{\frac{1}{2}}$$

$$=> k \cdot \sqrt{\frac{a}{2}} \approx \sqrt{\frac{1}{2}}$$

$$=> a = \frac{1}{k^2}$$

Thus, the optimized $reward\_factor$ ($a$) is given by:

$$reward\_factor\ (a) = \frac{(SR_1 - SR_2)^2}{n^2 \cdot SR_1 \cdot (100 - SR_1)}$$

Using above $reward\_factor$, the $score\_threshold$ can be expressed as:

$$score\_threshold = \frac{0.29 \cdot SR_1 + 0.71 \cdot SR_2}{100}$$

## 3.4 Practical Considerations

To account for system latency in receiving transaction responses, the score update method must satisfy:

$$\frac{SR_1}{100} \cdot (1 - a)^N > Threshold$$

where, $N = Avg\ TPS \cdot Avg\ Latency$

This ensures that scores degrade correctly even under delayed feedback conditions.

Additionally, by analysing the PG performance across multiple merchants, global downtime detection becomes even more significant, particularly for low-volume merchants who might otherwise struggle to identify such issues promptly. Integration with a payment orchestrator [1] enables us to leverage collective performance data across merchants, significantly enhancing the robustness and reliability of downtime detection systems.

## 3.5 Recovery and Reviving Gateways

After certain time, the system revives down gateways with a soft-reset score increment:

$$score\_new = score\_old \cdot (\frac{1}{(1-a)^{10}})$$

This allows gateways to process a limited number of transactions. If the underlying issue still persists, the gateway becomes down more rapidly, maintaining system stability.

# 4 Experimental Results

To assess the effectiveness of the implemented dynamic routing mechanism in a live production environment, JUSPAY conducted an in-depth analysis of its performance across multiple categories and dimensions. By comparing dynamic routing with traditional rule-based routing, the results demonstrate how the dynamic routing actually improves overall performance. The following sections present and discuss the outcomes, offering valuable insights into the system's performance and practical impact.

## 4.1 Rule-Based vs Dynamic Routing Success Rates for multiple Dimension

| Dimensions | % | Rule-Based Routing Success Rate | % | Dynamic Routing Sucess Rate | % | Traffic Split |
|---|---|---|---|---|---|---|
| ORDER_PAYMENT / UPI | | 78.42% | | 78.93% | | 46.13% |
| ORDER_PAYMENT / CARD | | 85.02% | | 86.08% | | 30.05% |
| ORDER_PAYMENT / NB | | 28.34% | | 30.87% | | 23.82% |

Figure 4 : Rule-Based vs Dynamic Routing across payment instruments

The table in figure 4 provides a high-level comparison of success rates between rule-based and dynamic routing for payment instruments: **UPI, Card** and **Net Banking (NB)** over **15** days. It also highlights the percentage of traffic each method contributes to the total system. The improvements in success rates vary across payment instruments, with each benefiting uniquely from dynamic routing. The cumulative improvement across all dimensions amounts to **1.15%**, reflecting a meaningful boost in the system's ability to process transactions successfully.

| Dimensions | % | Rule-Based Routing Success Rate | % | Dynamic Routing Sucess Rate |
|---|---|---|---|---|
| ORDER_PAYMENT / UPI / UPI_PAY | | 85.54% | | 85.87% |
| ORDER_PAYMENT / UPI / UPI_COLLECT | | 79.51% | | 81.72% |
| ORDER_PAYMENT / UPI / UPI_QR | | 82.78% | | 84.41% |
| ORDER_PAYMENT / CARD / DEBIT | | 65.76% | | 72.84% |
| ORDER_PAYMENT / CARD / CREDIT | | 87.17% | | 87.53% |
| ORDER_PAYMENT / CARD / CREDIT/VISA | | 68.85% | | 74.47% |

Figure 5 : Rule-Based vs Dynamic Routing across dimensions within UPI and CARD

The table in figure 5 demonstrates that dynamic routing consistently outperforms rule-based routing across various payment dimensions within **UPI** and **CARD** by dynamically adjusting decisions based on real-time success rates. This adaptability leads to significant improvements, with performance gains of up to **2%** in certain **UPI** dimensions and up to **7%** within **CARD**. Such enhancements showcase the system's flexibility and reliability in optimizing success rates, delivering better outcomes across diverse transaction scenarios and payment instruments.

## 4.2 Traffic Distribution among Gateways

| Gateways | % | Gateway Success Rate | % | Total Traffic Split Per Gateway |
|---|---|---|---|---|
| G1 | | 79.23% | | 65.60% |
| G2 | | 78.80% | | 20.10% |
| G3 | | 78.64% | | 14.40% |

Figure 6 : Traffic Distribution in Dynamic Routing

The traffic distribution in figure 6 shows how dynamic routing optimally allocates traffic based on gateway success rates. Gateway **G1**, with the highest success rate **(79.23%)**, handles the largest traffic share **(65.60%)**, while **G2** and **G3**, with slightly lower rates (**78.80%** and **78.64%**), receive **20.00%** and **14.40%**, respectively. This adaptive approach balances maximizing high-performing gateways' usage while maintaining real-time performance metric for all gateways, ensuring overall system efficiency.

## 4.3 Rule-Based vs Dynamic Routing Success Rates over time in UPI

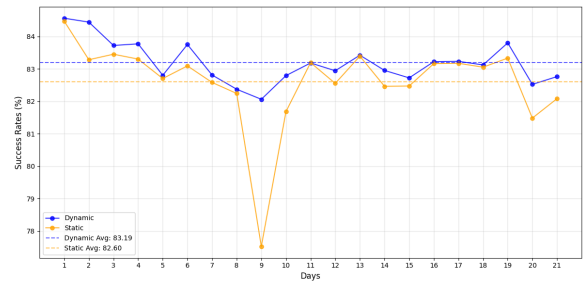

Figure 7 : Dynamic vs Rule-Based Success Rates over Days

The graph in figure 7 compares success rates of rule-based and dynamic routing over **21** days within UPI, highlighting dynamic routing's superior performance and stability. Dynamic routing maintains an average success rate of **83.19%**, adapting to real-time conditions to optimize performance, while rule-based routing averages **82.60%** and shows greater fluctuations. Notably, on day **9**, rule-based routing lags by **5%**, highlighting its reduced reliability. Dynamic routing's higher success rates and smaller variations ensure greater resilience and efficiency

## 4.4 Business Impact of Downtime Detection

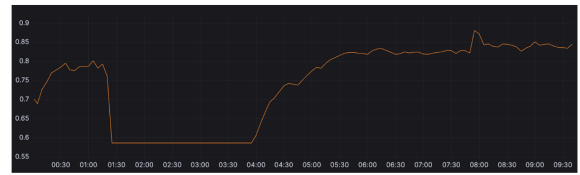

Figure 8 : Gateway Health score during Downtime instance

A significant downtime event began at **1:24 AM**, as shown in figure 8. The routing engine detected the issue within a minute, marked by a sharp drop in the gateway health score. Swift action rerouted **~8,000 transactions** to alternative gateways during the **2.5 hours downtime**, ensuring uninterrupted payment processing. This proactive response minimized disruptions and added **~$15,000** in **GMV** for a single merchant, demonstrating the system's capability to adapt to critical events and maintain business continuity.

# 5 CONCLUSION & FUTURE WORK

The proposed dynamic gateway routing system offers a robust and adaptable framework designed to optimize payment transaction success rates. By integrating real-time scoring, dynamic feedback loops and adaptive routing decisions, the system ensures fairness, reliability and scalability across various environments. Its exploration mechanisms help prevent gateway starvation, while the continuous update of performance data across multiple dimensions ensures consistent reliability. Furthermore, downtime detection strengthens system robustness by deprioritizing underperforming gateways based on merchant-defined metrics.

With JUSPAY serving as a leading payment orchestrator [1], this dynamic routing solution not only improves the technical performance of payment transactions but also creates more efficient routing decisions, ultimately enhancing both operational and financial outcomes. Looking to the future, the system will evolve with the addition of **ROI-based routing**, aligning gateway selection with the goal of merchant's cost reduction along with success rate. This will further optimize both performance and financial success for merchants.

The system has been modularized as an independent routing engine, available as an open-source implementation at https://github.com/juspay/decision-engine. This engine provides reusable decision and feedback modules that can be seamlessly integrated into any merchant's infrastructure. Ongoing developments build on this foundation to deliver greater efficiency, scalability, and adaptability in payment routing across varied business models.